\documentclass[twocolumn,secnumarabic,amssymb, nobibnotes, aps, prb]{revtex4-1}

\usepackage{graphicx}

\begin{document}


\title{Effect of \emph{in-situ} deposition of Mg adatoms on spin relaxation in graphene}



\author{Adrian G. Swartz}
\affiliation{Department of Physics and Astronomy, University of California, Riverside CA 92521}

\author{Jen-Ru Chen}
\affiliation{Department of Physics and Astronomy, University of California, Riverside CA 92521}

\author{Kathleen M. McCreary}
\affiliation{Department of Physics and Astronomy, University of California, Riverside CA 92521}

\author{Patrick M. Odenthal}
\affiliation{Department of Physics and Astronomy, University of California, Riverside CA 92521}

\author{Wei Han}
\affiliation{Department of Physics and Astronomy, University of California, Riverside CA 92521}

\author{Roland K. Kawakami}
\email[]{roland.kawakami@ucr.edu}
\affiliation{Department of Physics and Astronomy, University of California, Riverside CA 92521}


\date{\today}

\begin{abstract}
We have systematically introduced charged impurity scatterers in the form of Mg adsorbates to exfoliated single layer graphene and observe little variation of the spin relaxation times despite pronounced changes in the charge transport behavior. All measurements are performed on non-local graphene tunneling spin valves exposed \emph{in-situ} to Mg adatoms, thus systematically introducing atomic-scale charged impurity scattering. While charge transport properties exhibit decreased mobility and decreased momentum scattering times, the observed spin lifetimes are not significantly affected indicating that charged impurity scattering is inconsequential in the present regime of spin relaxation times ($\sim$1 ns).
\end{abstract}


\maketitle 

Graphene's gate tunable transport, tabletop relativistic physics, chemical attributes, and mechanical properties have interested researchers in a wide variety of fields.\cite{novoselov:2005,zhang:2005,castroneto:2009a,lee:2008} In particular, graphene is a candidate material for spintronics due to its weak hyperfine coupling and low intrinsic spin-orbit (SO) coupling strength ($\Delta_{SO}$),\cite{huertas:2006,min:2006,yao:2007,tombros:2007} which should theoretically lead to long spin lifetimes.  Beyond scientific interest, recent progress in large area production by chemical vapor deposition\cite{li:2009,bae:2010} combined with significant advances in efficient spin injection by improved tunneling contacts\cite{han:2010,dlubak:2012} has greatly improved the potential for advanced information processing utilizing spin-based logic.\cite{dery:2012} In particular, the introduction of efficient tunneling contacts has increased the observed spin lifetime by an order of magnitude (to a few ns in exfoliated graphene) by lengthening the escape time due to the backflow of electrons into the ferromagnetic leads.\cite{han:2010,yang:2011,han:2011} While graphene remains a highly promising candidate for carbon based spintronics, the observed spin lifetimes are still well below the theoretical expectations and the nature of spin relaxation remains an open question.

In graphene, two possible spin relaxation mechanisms are discussed in the literature:\cite{yang:2011,han:2011,ertler:2009,ochoa:2012,castroneto:2009b,zhang:2011,huertas:2007,huertas:2009,jozsa:2009,popinciuc:2009,zhang:2012} the Elliot-Yafet (EY) mechanism, for which the spin relaxation time ($\tau_s$) is proportional to the momentum scattering time ($\tau_p$), and the D'yakonov-Perel (DP) mechanism, for which $\tau_s \propto 1/\tau_p$. Complicating the situation are the many possible sources of spin relaxation in experiments on SiO$_2$ substrate including charged impurity (CI) scatterers,\cite{ertler:2009,ochoa:2012} Rashba SO coupling due toadatoms,\cite{castroneto:2009b,weeks:2011,zhang:2011} ripples,\cite{huertas:2007,huertas:2009} and edge effects.\cite{ochoa:2012,popinciuc:2009} Early experiments on spin transport in exfoliated graphene were able to take advantage of the tunable carrier concentration ($n$) and observe a linear relationship between $\tau_s$ and $\tau_p$, thus suggesting EY.\cite{jozsa:2009,han:2011} However, recent theoretical studies have shown that DP is expected to dominate over EY\cite{huertas:2009,zhang:2012} and that Elliot's approach applied to graphene\cite{ochoa:2012} predicts $\tau_s=(\epsilon_F)^2\tau_p/(\Delta_{SO})^2$, for which both Fermi energy $\epsilon_F$ and $\tau_p$ depend on carrier concentration, thus highlighting the need for experiments that can tune $\tau_p$ at fixed $n$.

In this work we systematically introduce CI scatterers on non-local single-layer graphene (SLG) spin valves with high quality tunneling contacts. The experiment takes place in an ultra-high vacuum (UHV) deposition chamber with \emph{in-situ} measurement capability at cryogenic temperatures. All measurements and doping are performed in the same chamber at $T$=12 K and the sample is never exposed to air. We choose Mg adsorbates as the CI scatterer since elements with low atomic weight should introduce minimal SO coupling. This substantially improves on earlier doping studies that utilized heavy atoms (Au) and ohmic contacts for shorter spin lifetimes($\tau_s \sim100$ ps)\cite{pi:2010}, which are dominated by contact induced spin relaxation.\cite{han:2010} We find that doping with Mg causes large shifts in the charge neutrality point (CNP), indicating significant charge transfer to the graphene layer, accompanied by increased momentum scattering. Spin transport measurements, however, indicate minimal effect on the spin relaxation, despite pronounced changes in charge transport. These results indicate that CI scattering is not an important source of spin relaxation in SLG in the current regime of spin lifetimes of $\sim$1 ns.

\begin{figure}
\includegraphics[width=85mm]{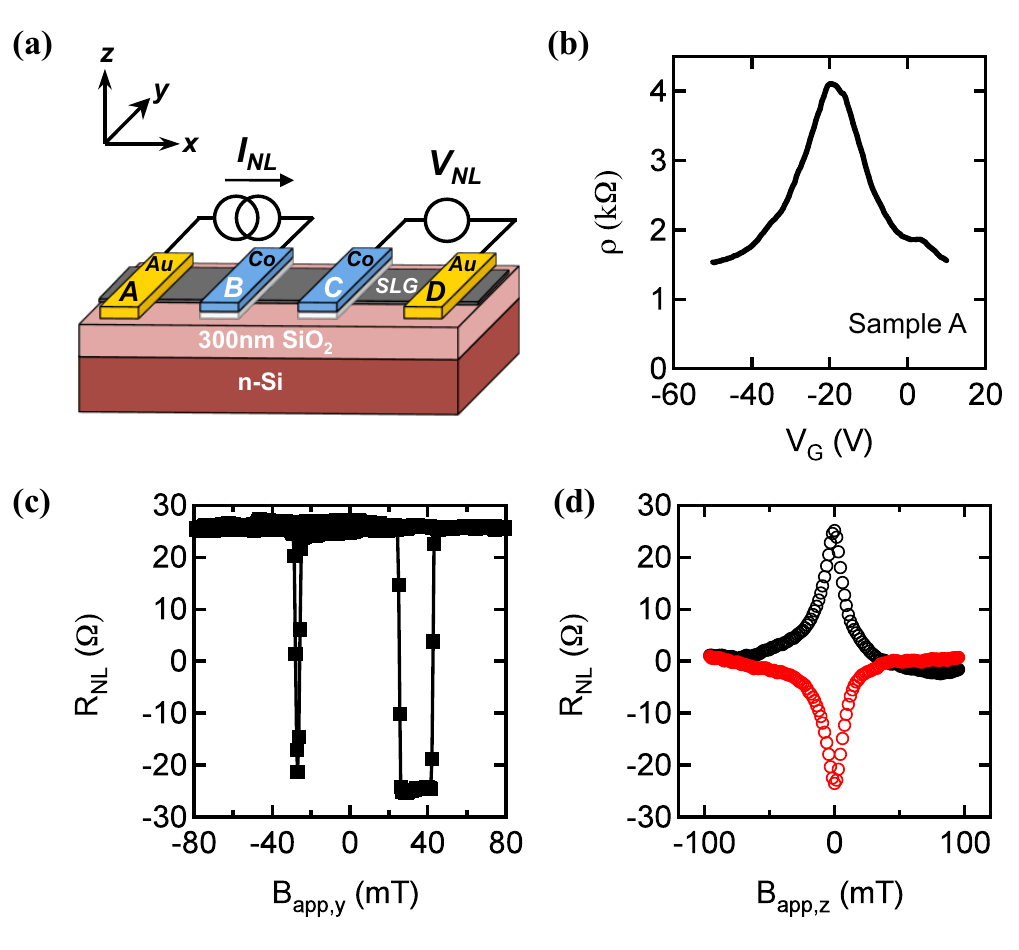}
\caption{\label{fig:pristine} (a), Device schematic of the non-local spin valve geometry with 
inner Co electrodes (blue)
and outer Au electrodes (yellow). (b) Gate dependent resistivity for 
sample A at $T=12$ K. (c) $R_{NL}$ for
pristine SLG at $V_G=0$ V. (d) Hanle spin precession data in 
parallel (black) and anti-parallel (red/grey)
configuration between electrodes B and C for pristine SLG at $V_G=0$ V. A constant spin-independent background has been subtracted.
}
\end{figure}

Graphene flakes are obtained by mechanical exfoliation of HOPG (SPI, ZYA) onto 300 nm SiO$_2$/Si. SLG flakes are identified under an optical microscope and confirmed by Raman spectroscopy. The graphene flakes are electrically contacted using standard bilayer (PMMA/MMA) e-beam lithography and lift-off procedures. First, outer Au/Ti electrodes (60 nm/8 nm) are defined and deposited by e-beam evaporation to serve as spin insensitive reference contacts. The sample is then annealed for 3 hours in UHV at 150$^\circ$C immediately prior to the second lithography step, which defines the inner ferromagnetic electrodes. Angle evaporation is utilized to deposit sub-monolayer TiO$_2$, which serves as a diffusion barrier for the 0.9 nm MgO tunnel barrier, and 80 nm Co. These tunneling contacts are deposited in a molecular beam epitaxy (MBE) chamber with base pressure of $1\times10^{-10}$ torr. The electrodes are then capped with 5 nm Al$_2$O$_3$. A detailed description for the fabrication of tunneling contacts is described elsewhere.\cite{han:2011}

Charge and spin transport measurements at $T$=12 K are performed on non-local devices as shown in Fig.~\ref{fig:pristine}a. The gate dependent resistivity of pristine SLG (sample A) is shown in Fig.~\ref{fig:pristine}b with maximum resistivity at the charge neutrality point, $V_{CNP}=-20$ V. The mobility is calculated by taking the slope of the conductivity ($\mu=\Delta\sigma/e\Delta n$) where the carrier concentration, $n$ (positive for holes), is determined using the relation $n=-\alpha(V_G-V_{CNP})$ and $\alpha=7.2\times10^{10}$ V$^{-1}$cm$^{-2}$ for 300 nm SiO$_2$ gate dielectric. The resulting electron and hole mobilities are $\mu_e=1774$ cm$^2$/Vs and $\mu_h=1508$ cm$^2$/Vs, respectively. For spin transport measurements, an AC current, $I_{NL}=1$ $\mu$A (11 Hz), is applied to inject spin-polarized carriers into SLG at electrode B. This spin polarization diffuses through the graphene channel along the $x$-axis to electrode C. A non-local voltage, $V_{NL}$, is detected using standard lock-in techniques between electrodes C and D due to the accumulation of spins beneath electrode C. The detected voltage, $V_{NL}$, is proportional to the spin-dependent chemical potential difference between electrodes C and D.\cite{tombros:2007} The non-local resistance, $R_{NL}=V_{NL}/I_{NL}$, depends on the relative orientation of the two inner ferromagnetic electrodes and is positive (negative) for parallel (antiparallel) alignment.  An external magnetic field, $B_{app,y}$, is applied along the electrode easy axis ($y$-axis) and is used to control the relative alignment of the magnetic electrodes. A typical sweep of $B_{app,y}$ for sample A at $V_G=0$ V ($n=-1.44\times10^{12}$ cm$^{-2}$) is shown in Figure \ref{fig:pristine}c, for which the spin signal $\Delta R_{NL}=R_{NL}^P-R_{NL}^{AP}$ is 50.5 $\Omega$. The dimensions of the graphene spin channel for sample A are defined by the channel length $L=2.2$ $\mu$m and width $w=2.4$ $\mu$m. The spin lifetime can be determined from Hanle spin precession measurements in which a magnetic field, $B_{app,z}$, is applied out of plane allowing the injected spins to precess around $B_{app,z}$. At large fields, the ensemble spin population dephases as $B_{app,z}$ is increased due to a distribution of arrival times at electrode C. In the tunneling limit, the ensemble spin precession can be fit using the Hanle equation,\cite{tombros:2007,han:2011}
\begin{equation}
R_{NL} \propto \int_{0}^{\infty} \frac{e^{-L^2/4Dt}}{\sqrt{4 \pi D t}}
\cos\! \left( \frac{g \mu_B B_{app,z}}{\hbar}\,t  \right) e^{-t/\tau_s} dt
\label{eq:hanle}
\end{equation}
where $D$ is the diffusion coefficient, $g$ is the electron $g$-factor, $\mu_B$ is the Bohr magneton, and $\hbar$ is the reduced Planck's constant. Figure \ref{fig:pristine}d shows characteristic Hanle curves for parallel and antiparallel alignment for $n=-1.44\times10^{12}$ $\text{cm}^{-2}$, where best fits to the Hanle equation yield the diffusion coefficient $D=0.058$ $\text{m}^2/\text{s}$, spin lifetime $\tau_s =1.10$ ns, and spin diffusion length $\lambda_s =8.0$ $\mu$m.

\begin{figure}
\includegraphics[width=80mm]{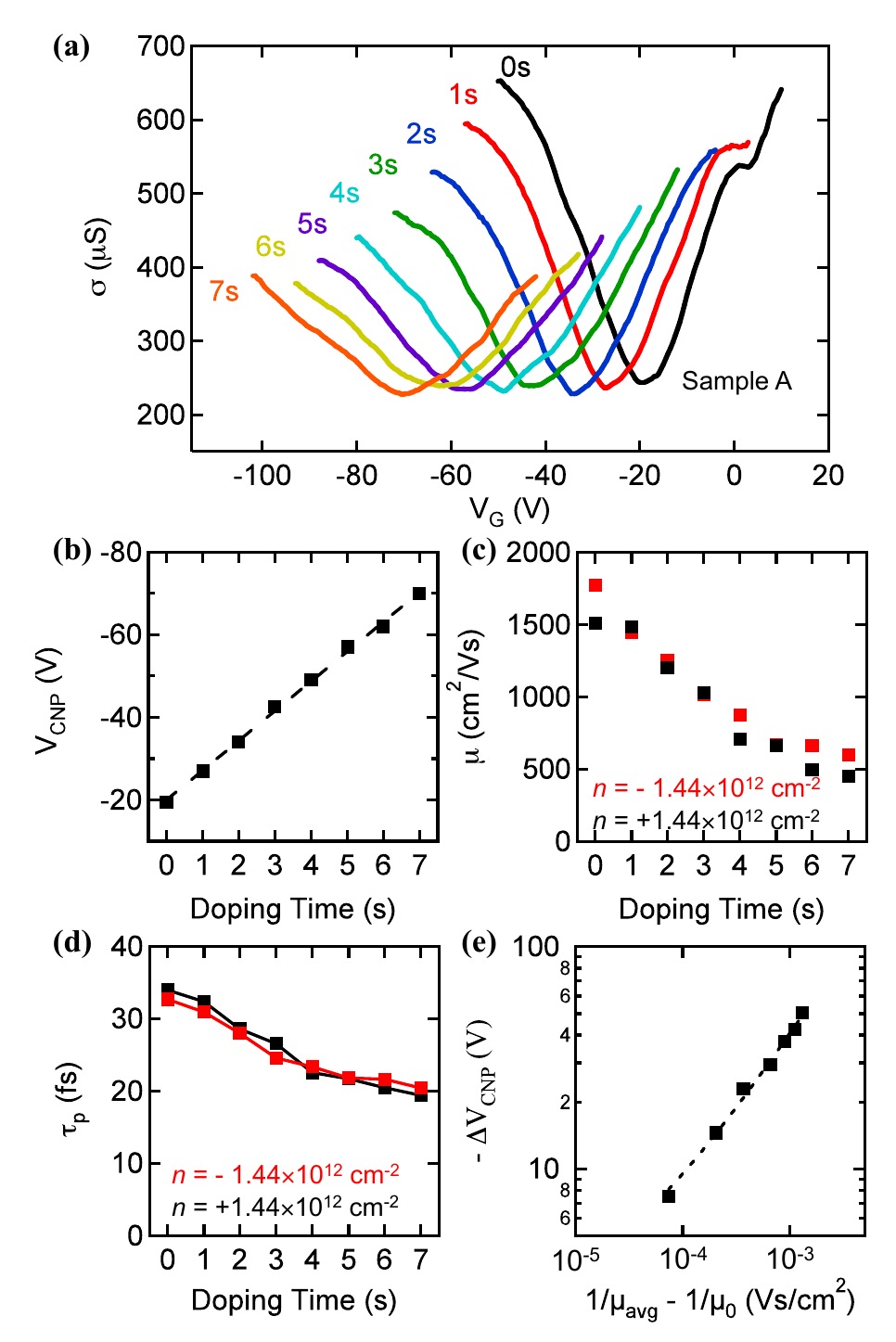}
\caption{\label{fig:charge} 
(a) Gate dependent conductivity as the SLG (sample A) is systematically doped up to 7 s
with Mg adsorbates at $T$=12 K. (b) Charge neutrality point (CNP) plotted against the Mg doping
time. (c) Electron (red/grey) and hole (black) mobility as a function of Mg coverage. (d) Calculated
momentum scattering time for electrons (red/grey) and holes (black) as a function of Mg coverage. (e)
Shift in the charge neutrality point, $-\Delta V_{CNP}$, plotted against the change in inverse mobility. The dashed line is a power law fit (best fit exponent $b=0.72$).
}
\end{figure}

Next, Mg adsorbates are deposited \emph{in-situ} in the UHV MBE chamber with base pressure $3 \times10^{-10}$ torr while the sample is maintained at $T$=12 K. Elemental Mg (99.99\%) is evaporated from an effusion cell at a rate of 0.055 \AA/min calibrated by a quartz crystal monitor and corresponds to a doping rate of 0.02\% of a monolayer (ML) per second,  where 1 ML is defined as $1.908\times10^{15}$ atoms/cm$^2$. After 1 s Mg deposition, the charge and spin transport are re-measured. Figure \ref{fig:charge} summarizes the effect on the charge transport on sample A following Mg doping. Figure \ref{fig:charge}a shows conductivity $\sigma$ vs. $V_G$ for Mg doping of sample A up to 7 s deposition time. After 7 s of Mg doping, $V_{CNP}$ has shifted to $V_G=-70$ V. This indicates that Mg donates electrons to the graphene, consistent with reports on transition metals and potassium.\cite{giovannetti:2008,pi:2009,mccreary:2010,chen:2008} Figure \ref{fig:charge}b displays $V_{CNP}$ for each doping time and demonstrates a linear relation between charge transfer and Mg coverage at a rate of $-1438$ V/ML.  Also, Mg doping introduces CI scattering which decreases the conductivity and the mobility. Figure \ref{fig:charge}c displays the effect of systematic Mg doping on the electron and hole mobilities. For undoped graphene, the mobility is $\mu_e=1774$ cm$^2$/Vs and and $\mu_h=1508$ cm$^2$/Vs, and decreases to $\mu_e=599$ cm$^2$/Vs and $\mu_h=453$cm$^2$/Vs after 7 s deposition time. The momentum scattering time can be determined using Boltzmann transport theory,\cite{tan:2007}
\begin{equation}
\tau_p=\frac{h \sigma}{e^2v_F\sqrt{n\pi g_s g_v}}
\label{eq:taup}
\end{equation}
where $h$ is Planck's constant, $e$ is the electron charge, $v_F\sim1\times10^6$ m/s is the Fermi velocity, and $g_e=2$ and $g_v=2$ are the spin and valley degeneracies. Fig.~\ref{fig:charge}d shows $\tau_p$ vs.~Mg doping for electrons and holes at $n=\pm1.44\times10^{12}$ cm$^{-2}$. With increasing Mg coverage, the momentum scattering time decreases due to increased CI scattering. Lastly, we investigate the nature of Mg morphology on the graphene surface. Figure \ref{fig:charge}e shows the shift in Dirac point plotted against $1/\mu_{avg}-1/\mu_0$, where $\mu_{avg}$ is the average of the electron and hole mobilities and $\mu_0$ is the average electron and hole mobility for pristine graphene. The dashed line is a power law fit of $-\Delta V_{CNP} \propto (1/\mu_{avg}- 1/\mu_0)^b$, for which values of $1.2<b<1.3$ indicates a $1/r$ scattering potential for point-like scatterers.\cite{chen:2008,adam:2007,pi:2009,mccreary:2010} The best fit value of $b=0.72$ suggests the possibility of clustering even at cryogenic temperatures.\cite{pi:2009,mccreary:2010} This does not introduce a theoretical difficulty because the relationship $\tau_s=(\epsilon_F)^2\tau_p/(\Delta_{SO})^2$ for EY scattering in SLG has been shown to hold for a wide variety of scattering sources including weak scatterers, strong scatterers (i.e. vacancies), CI scatterers, and clusters.\cite{ochoa:2012} Lastly, we note that the gate dependent resistance curves exhibited no measurable change as a function of time in between Mg depositions. 

\begin{figure}
\includegraphics[width=85mm]{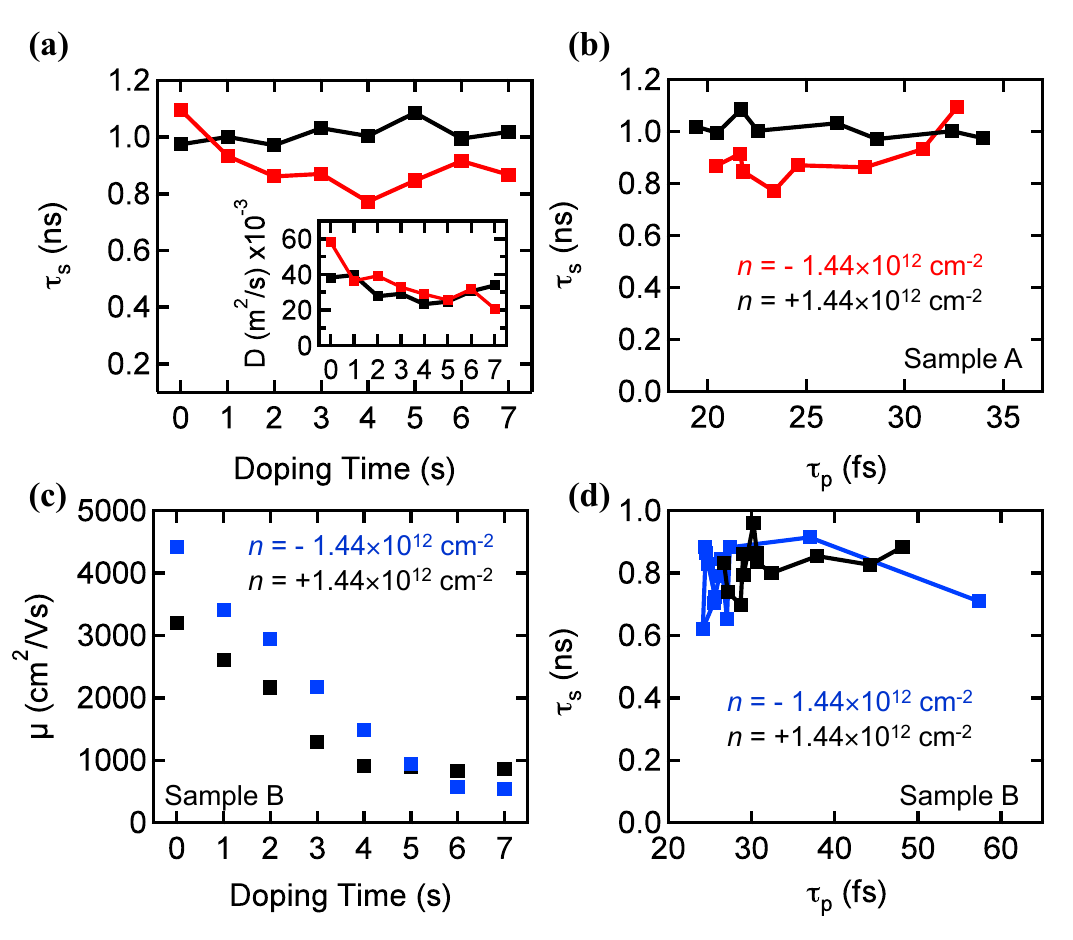}
\caption{\label{fig:spin} 
(a) Hanle spin lifetime for sample A plotted against Mg coverage for electrons (red/grey) and holes (black). Inset is the diffusion
coefficient obtained from the Hanle fit as a function of Mg coverage at $T$=12 K. (b) Hanle spin
lifetime for sample A plotted against the calculated momentum scattering times for each doping for both electrons (red/grey) and holes (black).
(c) Electron (blue/grey) and hole (black) mobility as a function of Mg coverage for sample 
B at $T$=12 K. (d) Hanle spin lifetime for sample B plotted
against the calculated momentum scattering times.
}
\end{figure}

We now turn to the effect on spin relaxation in SLG by Mg doping. After each Mg deposition at 1 s intervals, Hanle spin precession measurements were performed for $n=\pm1.44\times10^{12}$ cm$^{-2}$. The resulting fits to the Hanle curves yield values for $\tau_s$ and $D$ which are plotted against Mg doping time in Figure \ref{fig:spin}a and \ref{fig:spin}a inset, respectively. The diffusion coefficient decreases with increasing Mg coverage starting at 0.058 m$^2$/s (0.038 m$^2$/s) for pristine graphene and decreases to 0.021 m$^2$/s (0.033 m$^2$/s) for 7 s doping time for electrons (holes). This is in agreement with the observed charge transport behavior for which momentum scattering increases with Mg doping. Interestingly, the spin lifetimes (Fig.~\ref{fig:spin}a) show minimal variation, without a significant trend for electrons and holes. In Figure \ref{fig:spin}b we plot the Hanle spin lifetime for sample A  against the momentum scattering time calculated from the conductivity using equation (\ref{eq:taup}) from Boltzmann transport theory for sample A. With increasing Mg doping, $\tau_p$ decreases from $\sim35$ fs to $\sim20$ fs, but the spin relaxation time is constant for holes (black squares) while decreasing only slightly for electrons (red/grey squares). This experiment was repeated on several samples and in general $\tau_s$ does not display any substantial variations as a function of $\tau_p$. For instance, results for a sample with higher initial mobility (sample B) are summarized in Fig.~\ref{fig:spin}c and 3d.  Figure \ref{fig:spin}c displays the change in mobility for electrons and holes under Mg doping. For sample B, the mobility decreases from 4415 cm$^2$/Vs (3200 cm$^2$/Vs) for the pristine spin valve to 598 cm$^2$/Vs (1290 cm$^2$/Vs) after 7 s Mg doping for electrons (holes). In Fig.~\ref{fig:spin}d, we show $\tau_s$ displayed against the momentum scattering times for sample B at $n=\pm1.44\times10^{12}$ cm$^{-2}$. Here, $\tau_s$ is near 800 ps and stays relatively unchanged as $\tau_p$ decreases from $\sim60$ fs to $\sim24$ fs.

As Fig.~\ref{fig:spin}b and 3d show, $\tau_s$ does not vary substantially as $\tau_p$ is varied by CI scattering. This is in agreement with and goes beyond recent reports on CI scattering by organic-ligand bound nanoparticles, which are able to reversibly tune the mobility and momentum scattering.\cite{han:2012} Due to the relatively large size ($\sim13$ nm, which is over 50 lattice constants) of the nanoparticles used in that study, it is not possible to draw conclusions for atomic-scale CI scatterers such as surface adatoms and impurities in the SiO$_2$ substrate. In contrast, Mg adsorbates are able to probe the atomic-scale regime. With the agreement between two quite different experiments (Mg adsorbates deposited in UHV and organic-ligand bound nanoparticles deposited by drop casting) probing different length scales of the scattering potential, it is clear that spin relaxation in graphene is not determined by CI scattering despite its importance for momentum scattering. 

It is also worth mentioning that this result is not incompatible with the early experiments showing a linear relation between $\tau_s$ and $D$ by tuning the carrier concentration.\cite{jozsa:2009,han:2011} While Mg adsorbates modify $\tau_p$ by introducing CI scattering and possibly local Rashba SO coupling, there are many alternative sources which might contribute to EY (i.e. weak scatterers, resonant scattering, phonon scattering) which could still present themselves as the carrier concentration is tuned leading to $\tau_s\sim D$. Thus, EY spin relaxation originating from sources other than CI scattering is still viable.

Some other possibilities to consider are DP spin relaxation in spatially inhomogeneous Rashba SO fields. It has recently been proposed that this type of SO coupling can result in a competition between EY-like and DP-like scaling behavior to yield unconventional scaling between $\tau_s$ and $\tau_p$.\cite{zhang:2011} Another possibility is that the spin lifetime is limited by contact effects such as inhomogeneous stray fields.\cite{dash:2011} Due to its atomically thin nature, this could have a larger effect for graphene compared to semiconductor or metallic spin transport systems that are typically much thicker.

In conclusion, we have investigated charge and spin transport in SLG by systematically introducing CI scatterers in the form of Mg adsorbates. The introduction of Mg was shown to transfer electrons to the SLG and decrease the momentum scattering time. Despite pronounced changes in momentum scattering, no significant variation was seen in spin relaxation. This indicates that CI scattering is not an important source of spin relaxation in SLG in the current regime of spin lifetimes ($\sim$1 ns).

We acknowledge support from NSF (DMR-1007057, MRSEC DMR-0820414), ONR (N00014-12-1-0469), and NRI-NSF (NEB-1124601).


%

\end{document}